\documentstyle[fleqn,11pt]{article}
\title{
$K^*(892)$ Electromagnetic Mass Anomaly
\thanks{Talk presented in the BES annual conference, 2000, Dalian,
China.
 The work was supported in part by National Fund of China.}}
\author{
{Mu-Lin Yan\thanks{E-mail address:mlyan@ustc.edu.cn}}
\\
Center for Fundamental Physics \\
University of Science and Technology of China\\Hefei Anhui 230026, P.R.China\\}
\date{Dec. 7, 2000}

\begin{document}
\maketitle
\begin{abstract}
{Electromagnetic masses of neutral $K^*(892)$ may be larger than
one of charged $K^*(892)$. It is unusual and is called $K^*$-EM-mass
anomaly. We review the studies on this issue, and point out that
$K^*$-mass splitting can be measured in BES accurately. }
\end{abstract}
\begin{enumerate}
\item {\bf $K^*(892)$}: The isospin, spin and parity of meson resonance $K^*(892)$
are $I(J^P)=1/2(1^-)$. Its flavor SU(3) multiplets are
$\bar{K}^{*0}=\bar{d}\gamma_\mu s$, ${K}^{*0}=\bar{s}\gamma_\mu
d$, ${K}^{*+}=\bar{s}\gamma_\mu u$ and ${K}^{*-}=\bar{u}\gamma_\mu
s$. Among the meson-resonances, $K^*(892)$ is only one whose mass
difference between its isospin components has been determined in
experiments although the error bar is still large. According to
the 2000-PDG report, $K^*$ mass splitting is\cite{pdg}
\begin{equation}
m_{K^*(892)^0}-m_{K^*(892)^\pm}=6.7\pm1.2 MeV.
\end{equation}
Note, this value is larger than the mass splitting between $K^0$
and $K^\pm$,
\begin{equation}
m_{K^0}-m_{K^\pm}=3.995\pm0.034 MeV.
\end{equation}
\item {\bf Mass splitting of hadrons}: Generally, for the SU(3)
flavor multiplets of hadrons, the mass splittings between their
isospin components are caused by two effects: i) $m_u\neq m_d$
(inequality of u- and d-quark masses); ii) the electromagnetic
interactions inside hadrons. Then, the observing quantity of the
splitting $(\Delta m)_{expt}$ is expressed as follows
\begin{equation}
(\Delta m)_{expt}=(\Delta m)_{Q}+(\Delta m)_{EM}
\end{equation}
where $(\Delta m)_{expt}$ is the experimental value; $(\Delta
m)_Q$, the contributions due to u-d quark mass difference, and
$(\Delta m)_{EM}$ is due to electromagnetic interactions, or
EM-mass difference. The theoretical predictions of $(\Delta m)_Q$
and $(\Delta m)_{EM}$ are model-dependent. They reflect the
effects of internal structures of hadrons and the hadron dynamics.

\item {\bf EM-mass anomaly}: Usually, the EM-masses of neutral
hadrons are less than one of their charged partners. For
instance, the EM-masses of neutron, $\pi^0$ and $K^0 (\bar
{k}^0)$ are less than the EM-masses of proton, $\pi^\pm$ and
$K^\pm$ respectively \cite{lyl}\cite{das}\cite{gao}. However,
some theoretical studies show that the EM-mass of neutral
$K^*(892)$ may be larger than the EM-mass of charged $K^*(892)$.
This is very unusual. In ref.\cite{yan}, we called it as
$K^*(892)$ EM-mass anomaly.

\item {\bf Estimation of EM-mass splitting of $K^*(892)$:}
Theoretically, $(\Delta m)_Q$ can be calculated in some models. By
using eq.(3) and experimental data $(\Delta m)_{expt}$, then, one
can predict  $(\Delta m)_{EM}$. There are two studies to estimate
$(\Delta m)_{Q}$ for $K^*$ in the literature:
\begin{enumerate}
  \item In ref.\cite{sch}, Schechter et al use an effective theory
  of hadrons to get $(\Delta m)_{Q}$ is in the region from
  $2.04MeV$ to $6.78MeV$. And its best parameter fitting result is
  \begin{equation}
  (m_{K^{*0}}-m_{K^{*\pm}})_Q=4.47MeV.
  \end{equation}
  This estimation indicates that so long as
  $(m_{K^{*0}}-m_{K^{*\pm}})_{expt}>6.78MeV$, the $K^*$-EM-anomaly
  exist surely. And under the best parameter fitting, the
  condition for the anomaly becomes
  $(m_{K^{*0}}-m_{K^{*\pm}})_{expt}>4.47MeV$. Consequently,
  from eq.(1), it can be seen that the experiment favors to
  support $(m_{K^{*0}}-m_{K^{*\pm}})_{EM}>0$, i.e., there exist
  $K^*$-EM-mass anomaly. But, because the error bar is still
  large, more precise experimental measurement for $K^*$ mass
  splitting is necessary.
  \item Using Li's chiral quark model\cite{li}, Gao and Yan found
  out\cite{gy} that
  \begin{equation}
  (m_{K^{*0}}-m_{K^{*\pm}})_Q \simeq {1\over 2} (m_d-m_u).
  \end{equation}
  From $\rho^0-\omega$ mixing, they found out further
  \begin{equation}
  m_d-m_u=6.14\pm0.36MeV.
  \end{equation}
  Thus, they predict
  \begin{equation}
  (m_{K^{*0}}-m_{K^{*\pm}})_Q=3.07\pm0.18MeV.
  \end{equation}
  Consequently, so long as
  $(m_{K^{*0}}-m_{K^{*\pm}})_{expt}>3.25MeV$, the $K^*$-EM-mass
  anomaly exists. Comparing with eq.(1), we see that the present
  data support there exist such an anomaly in Li's chiral quark
  model.
  \end{enumerate}
\item {\bf Significance:} $K^*$s are composite particles with
quark structures. Neutral particle $K^{*0}(892)$ is composited of
positive charged $\bar{s}-$quark and negative charged $d-$quark.
Due to the interactions between quarks, the distribution of the
positive charge "quark cloud" inside $K^{*0}$ is generally
different from one of the negative charge "quark cloud". This is
the reason why the neutral particle $K^{*0}$ is possessed of
EM-mass (or EM-self energies). As the distribution of the
positive charge cloud just overlaps one of the negative's, its
EM-mass would vanish. As the overlapping is few, or they are
totally dis-overlapped, the EM-mass of $K^{*0}$ will become so
large that it is bigger than the EM-mass of $K^{*\pm}$. This is
the case of $K^*-$EM-mass anomaly. Obviously, this anomaly effect
reflects important structure information on the quark
distribution inside vector meson resonance. Hence, it is valuable
for hadron physics and for QCD at middle energies to study this
topic experimentally and theoretically.
\item {\bf Principle to measure $K^*(892)$ mass splitting in BES
at BEPC:} BES at BEPC has gathered  about $5\times 10^7 J/\Psi$
events. Therefore, it is practicable to take $J/\Psi$ as the
source of $K^*(892)$ and to study its mass-splitting accurately.
The branch ratio for $J/\Psi\longrightarrow K^{\bar{+}}K^{* \pm}$
is $5\times 10^{-3}$; for $J/\Psi\longrightarrow
\bar{K}^0K^{*0}$, it is $4.2\times 10^{-3}$; for
$J/\Psi\longrightarrow K\bar{K}\pi$, $6\times 10^{-3}$; for
$K^*\longrightarrow K\pi$, it is about $100\%$\cite{pdg}.
Therefore, by using three-body decay processes of
$J/\Psi\longrightarrow \bar{K}K\pi$, we can determine the
locations of resonance $K^{\bar{+}}\pi^{\pm}$($=K^{*0}$ or
$\bar{K}^{*0}$) and resonance $K^{0}\pi^{\pm}$($=K^{*\pm})$
respectively. The event number of $J/\Psi\longrightarrow
\bar{K}K\pi$ in BES is about $10^5$. It is expected to decrease
the error for the splitting to $1MeV$ or below $1MeV$. This
should be a meaningful experiment in hadron physics. As the
apparatuses of energy measure in BES are improved further in the
near future, one could expect more accurate results on this
object.
\end{enumerate}

\end{document}